\documentclass{PoS}

\newcommand{\ba}{\begin{eqnarray}}
\newcommand{\ea}{\end{eqnarray}}
\newcommand{\bas}{\begin{eqnarray*}}
\newcommand{\eas}{\end{eqnarray*}}
\newcommand{\be}{\begin{equation}}
\newcommand{\ee}{\end{equation}}
\newcommand{\bes}{\begin{equation*}}
\newcommand{\ees}{\end{equation*}}
\newcommand{\bi}{\begin{itemize}}
\newcommand{\ei}{\end{itemize}}

\newcommand{\bcentre}{\begin{center}}
\newcommand{\ecentre}{\end{center}}

\font\tenmsb=msbm10 scaled\magstep1
\font\sevenmsb=msbm7 scaled\magstep1
\font\fivemsb=msbm5 scaled\magstep1
\newfam\msbfam
\textfont\msbfam=\tenmsb
\scriptfont\msbfam=\sevenmsb
\scriptscriptfont\msbfam=\fivemsb

\newcommand{\order}[1]{{\mathcal O}(#1)}

\newcommand{\mres}{m_{\rm res}}
\usepackage{epsfig}

\title{Baryons in 2+1 flavour domain wall QCD}
\author{
D.J.~Antonio, K.C.~Bowler, P.A.~Boyle, M.A.~Clark, B.~Jo\'o,
A.D.~Kennedy, R.D.~Kenway, \speaker{C.M.~Maynard}, R.J.~Tweedie.
	\\School of Physics, University of Edinburgh, Edinburgh, EH9 3JZ, UK.\\ 
	E-mail: \\ \email{$\ \ $s0459477@sms.ed.ac.uk}, 
		\\ \email{$\ \ $kcb@ph.ed.ac.uk}, 
		\\ \email{$\ \ $paboyle@ph.ed.ac.uk},
		\\ \email{$\ \ $mike@ph.ed.ac.uk},
		\\ \email{$\ \ $bj@ph.ed.ac.uk},
		\\ \email{$\ \ $adk@ph.ed.ac.uk},
		\\ \email{$\ \ $r.d.kenway@ed.ac.uk},
		\\ \email{$\ \ $cmaynard@ph.ed.ac.uk},
		\\ \email{$\ \ $rjt@ph.ed.ac.uk}}
\author{A.~Yamaguchi\\Department of Physics and Astronomy, University of Glasgow, Glasgow, G12 8QQ, UK.\\
		E-mail: \\\email{$\ \ $a.yamaguchi@physics.gla.ac.uk}}
\author{RBC and UKQCD Collaborations}
\abstract{
We present results for some of the light baryon masses and their
excited states in 2+1 flavour domain wall QCD. We considered
several lattice spacings, with the DBW2 and Iwasaki gauge actions and
different sea quark masses on a volume of $16^3\times32$ and a fifth
dimension of size 8.  All data were generated on the QCDOC machines.
Despite large residual massses and a limited number of sea quark mass
values with which to perform chiral extrapolations, our results are in
reasonable agreement with experiment and scale within errors. Finite size
effects on most ensembles appear to be small.
}

\PoS{PoS(LAT2005)098}
\ShortTitle{Baryons in 2+1 flavour domain wall QCD}
\FullConference{XXIIIrd International Symposium on Lattice Field Theory
\\Trinity College, Dublin, Ireland\\25-30 July 2005}
\begin{document}

\section{Introduction}
The calculation of many quantities from lattice QCD with quenched, or
heavy dynamical, quarks has produced results which, although in
qualitative agreement with experiment, have uncontrolled sources of
error. To progress beyond this, calculations with light dynamical
quarks are necessary. Using fermions which satisfy the Ginsparg-Wilson
(GW) relation is advantageous, as the theory is well defined within
broad parameter ranges. It has lattice versions of the symmetries of
continuum QCD, and the correct flavour content. Mixing of operators
with different chirality is suppressed, renormalisation is simplified,
and a continuum-like chiral perturbation theory can be used for the
extrapolation of quantities to the chiral limit. Domain wall fermions
(DWF) satisfy these requirements. The QCDOC machine and the RHMC
algorithm have allowed for the first time calculations using DWF with
$2+1$ flavours of dynamical quarks.

Baryon physics is an area rich in
phenomenology. Unanswered questions as to the nature of some of the
excited states, the decay of the proton predicted by some Grand
Unified and super~symmetric models, and the determination of 
matrix elements related to the structure functions and the neutron electric
dipole moment can, in principle, be determined by lattice QCD
calculations. These calculations are very computationally challenging,
as they need both light sea quarks and large volumes, as well as fermions
with the correct symmetry properties. In this work we report on a study
of the lowest lying states in the baryon spectrum, $\{N , \Delta ,
\Omega , N^{\star}\}$, on ensembles of configurations produced by
QCDOC. It is important for any calculation claiming to be ``full QCD''
to be able to reproduce this spectrum.  These initial ensembles were
produced primarily for a search of parameter space to guide larger
production runs.  As such, they have a small volume, too small
certainly for excited states such as the $N^{\star}$, but this can be
used, eventually in combination with the productions runs on larger
volumes to try and estimate the size of the finite volume effects on
the remaining spectrum.

The DWF action is given in~\cite{Furman:1994ky} with
the Pauli-Villars field in~\cite{Vranas:1997da} for the dynamical
simulation. The gauge fields were generated with renormalisation group
(RG) improved actions, as follows:
\begin{equation}
  S_G=-\frac{\beta}{3}\left[ (1-8) c_1 \sum_{x,\mu\nu} P(x)_{\mu\nu} +
  c_1 \sum_{x,\mu\neq\nu} R(x)_{\mu\nu}\right]
\end{equation}
with either $c_1=-1.4069$ for the DBW2 action~\cite{deForcrand:1999bi}
or $c_1=-0.331$ for the Iwasaki action~\cite{Iwasaki:1984cj}. It has
been noted in the quenched DWF calculations~\cite{Blum:2000kn} that
these actions reduce the chiral symmetry breaking resulting from the
finite fifth dimension. The mechanism and size of chiral symmetry
breaking on these ensembles is studied in detail in~\cite{PeterBoyle}.

In the DWF formalism, the 4D quark fields are constructed
from left and right projections of the 5D fermion fields
on the boundaries. With a finite fifth dimension, there is still
an overlap between these left- and right-handed fields, which manifests
itself as an additive mass renormalisation, known as the residual mass.
This can be determined directly from the axial Ward-Takahashi identity
\ba
  \partial_\mu A_\mu(x)& = & 2am_fP(x)+ 2 J_5(x)\\\nonumber
               & \approx & 2 (am_f + am_{\rm res})P(x)
\ea
where $J_5(x)$ is the point-split current constructed from fields at the 
mid-point of the fifth dimension and $P(x)$ the pseudoscalar density.

\section{Details of the calculation}
The ensembles were all generated on QCDOC machines, using the exact
RHMC algorithm~\cite{Kennedy:1998cu,MikeClark}, with a volume of
$16^3\times32$ and a fifth dimension of size $L_S=8$ and are detailed in
Table~\ref{tab:Ensembles}. The integrated autocorrelation times are
estimated to be $\order{50}$ for mesonic correlators. To maximise
statistics, the correlation functions were measured every 5
trajectories, and then binned so that the separation between
independent measurements is 100 trajectories. The correlators were
computed from up to four sources on different time-planes with as many as
three different smeared sources.

\begin{table}
\begin{center}
\begin{tabular}{cccccccc}
$\beta$ & $am_l/am_s$ & $r_0/a$ & $a^{-1}$(GeV) & $L$(fm) & $Lm_P$ & $m_P/m_V$ & $\#$traj \\\hline
$0.72$ & 0.04/0.04 & & & & $7.7(1)$ & $0.692(5)$ & 3400\\
$0.72$ & 0.02/0.04 & \raisebox{1.5ex}[0pt]{$4.3(1)$} &\raisebox{1.5ex}[0pt]{$1.7(1)$} & \raisebox{1.5ex}[0pt]{$1.9(1)$} & $6.0(1)$ & $0.589(3)$ & 6000 \\\hline
$0.764$ & 0.04/0.04 & & & &$6.7(1)$ & $0.699(4)$ & 5750\\
$0.764$ & 0.02/0.04 & \raisebox{1.5ex}[0pt]{$5.1(2)$} &\raisebox{1.5ex}[0pt]{$2.0(1)$} & \raisebox{1.5ex}[0pt]{$1.6(1)$} & $5.1(1)$ & $0.619(4)$ & 3000 \\\hline
$2.13$ & 0.04/0.04 & & & & $7.5(1)$ & $0.700(8)$ & 3600\\
$2.13$ & 0.02/0.04 & \raisebox{1.5ex}[0pt]{$4.6(2)$} &\raisebox{1.5ex}[0pt]{$1.8(1)$} & \raisebox{1.5ex}[0pt]{$1.8(1)$} & $5.8(1)$ & $0.615(5)$ & 3600 \\\hline
$2.2$ & 0.04/0.04 & & & & $6.8(1)$ & $0.726(2)$ & 4500\\
$2.2$ & 0.02/0.04 & \raisebox{1.5ex}[0pt]{$5.3(1)$} &\raisebox{1.5ex}[0pt]{$2.1(1)$} & \raisebox{1.5ex}[0pt]{$1.5(1)$}& $5.1(1)$ & $0.667(8)$ & 3200 \\\hline
\end{tabular}
\caption{Properties of the ensembles used in this study. The value of $r_0/a$ was determined in~\cite{KoichiHashimoto}.
The lattice spacing, and thus the volume are set by choosing $r_0=0.5$fm. Trajectories are of length 0.5}
\label{tab:Ensembles}
\end{center}
\end{table}

The masses of the pseudoscalar and vector mesons were determined by simultaneous
fits to the local and smeared correlators with two exponentials, the ground state and the
excited state. Similarly $\mres$ was determined by a simultaneous fit of a plateau to 
a ratio of correlators, with both smeared and local sources.

\begin{figure}
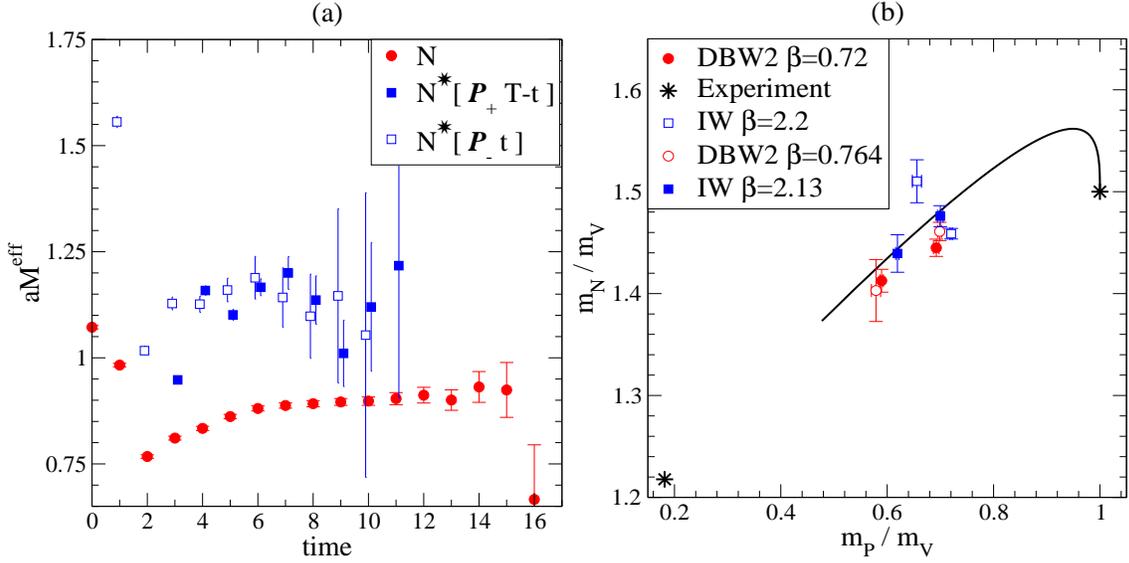

\begin{center}
  \epsfig{file=Nstar_m0.02.eps,height=0.49\textwidth,width=0.49\textwidth}
  \epsfig{file=EdinburghPlot.eps,height=0.49\textwidth,width=0.49\textwidth}
  \caption{(a) The effective mass of $I=\frac{1}{2}$ baryon correlator on the DBW2
    $\beta=0.72$, $am_l/am_s=0.02/0.04$ ensemble. The red symbol shows
    the data for the positive parity state, the closed blue squares are
    the time-reversed backward propagating negative parity state data and
    the open blue squares are the negative parity data from the second correlator.
    (b) The Edinburgh plot for the different ensembles.}
  \label{fig:effMass}
\end{center}
\end{figure}

The standard baryon interpolating operator is given by
\be
  \Omega(x)=\epsilon_{ijk}\left[\psi_i(x) C \Gamma \psi_j(x)\right]\psi_k(x)
\ee
For the $I=\{\frac{1}{2},\frac{3}{2}\}$ baryons,
$\Gamma=\{\gamma_5,\gamma_5\gamma_k\}$.  Another operator, which
projects onto the negative parity $I=\frac{1}{2}$ state, with
$\Gamma=1$ was also used.  For baryon correlators in a finite box
with periodic boundary conditions, the backward propagating state is
the negative parity partner, that is
\be
  C_B(t) = A_+e^{-m_+t} + A_-e^{-m_-(T-t)}
  \label{eqn:baryonCorr}
\ee
For the $I=\frac{1}{2}$ baryon, the masses of the positive and negative
parity states were determined by a simultaneous fit to equation
(\ref{eqn:baryonCorr}) using the standard operator, and a single
exponential to the negative parity correlator. This is shown in
Figure~\ref{fig:effMass}(a).  Typically this was computed for the
smeared correlator only, as the local correlator had a poor
signal. 

The mass of the $I=\frac{3}{2}$ baryon was determined from a single
exponential fit to the smeared correlator, as the signal for the
excited, negative parity partner was poor, as might be expected on
relatively small numbers of configurations.

\section{Preliminary Results}
Shown in Figure~\ref{fig:effMass}(b) is the Edinburgh plot. It is
reassuring that, even at relatively coarse lattice spacing, with a
small fifth dimension and consequently moderate chiral
symmetry breaking, the data follows the phenomenological curve very well. 
The only exception is the lightest Iwasaki $\beta=2.2$ datum. Naively, one
might expect this to be a finite size effect, especially, examining 
Table~\ref{tab:Ensembles}, given that the lattice spacing
and thus the box size are rather small. However, the value of $Lm_P$ is not
significantly smaller than the other data sets and, critically, $Lm_P>4$, suggesting
that the box is big enough as measured by the pseudoscalar meson. 

For this particular ensemble the signal for the vector meson mass is
not good. In particular, the effective mass plot has a poor plateau,
and a stable fit can only be achieved for the lowest region. The net
effect is for a rather low vector meson mass. It is probable that this
is due simply to low statistics. A low estimate of the vector mass
would cause this datum in the Edinburgh plot to be shifted up and to the
right, which could be mistaken for a finite volume effect increasing
the mass of the nucleon.

The quark mass was defined as
\be
  am_q=am_f + am_{\rm res}(m_f)
\ee
with the chiral limit at $am_q=0$. With two sea quark masses,
only a crude chiral extrapolation could be attempted, {\em i.e.} drawing a
straight line through the two data points. The strange quark mass, (for the
$\Omega$ baryon) was set from the kaon mass.

\begin{figure}
\begin{center}
  \epsfig{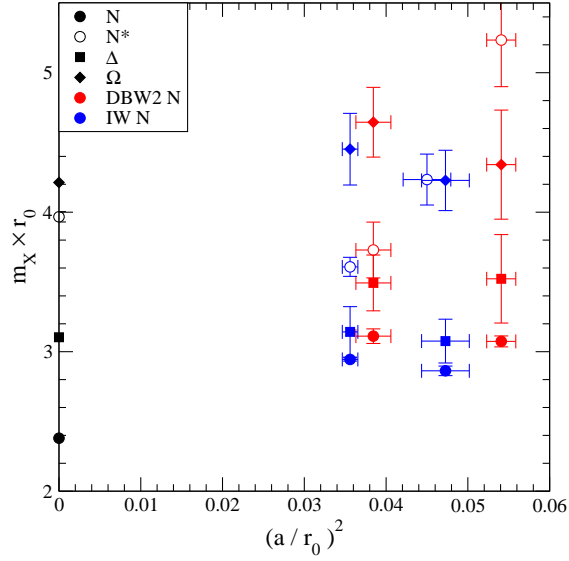}
  \caption{Scaling of the baryon spectrum with lattice spacing squared.
	   The symbols denote the following: Closed circles $N$, open circles
	   $N^{\star}$, squares $\Delta$, diamonds $\Omega$. Black symbols denote
           experiment, red DBW2 ensembles, blue Iwasaki ensembles.
	   The value of $r_0=0.5$fm was chosen, to give an indication of 
	   the experimental spectrum in these units. The $N^{\star}$ measured
	   on the $\beta=2.13$ ensemble (the furthest right open blue circle) has
	   been offset to the left for clarity.}
  \label{fig:Scaling}
\end{center}
\end{figure}

Shown in Figure~\ref{fig:Scaling} is the dependence of the baryon
spectrum, in dimensionless units, on the lattice spacing. A continuum
extrapolation cannot be attempted with these ensembles. However, the
data for the ground states, $\{N,\Delta ,\Omega\}$, shows reasonable
scaling, albeit with large errors.  These large uncertainties are due
to the crude nature of the chiral extrapolation. The determination of
$r_0$ is well defined for each ensemble, and thus it is a good
quantity to use to examine scaling behaviour. As the value of $r_0$ in
the continuum is unknown, rather than setting the absolute scale with
$r_0$, a better strategy is to predict dimensionless ratios of
physical quantities in the continuum, while using $r_0$ to just
examine the scaling behaviour.

The negative parity partner of the nucleon, the $N^\star$ is expected
to become degenerate with the nucleon in a small enough box. This
effect can be clearly seen from Figure~\ref{fig:Scaling}; the
$N^\star$ mass drops dramatically as the volume is reduced. This
suggests that finite size effects may also be beginning to affect the
ground states for the ensembles at finest lattice spacing. These
finite size effects would tend increase the mass of the ground
states. The slight upward tendancy in the scaling plot as the lattice
spacing decreases is consistent with the finite size effects spoiling
otherwise very good scaling, or a small scaling violation for the
nucleon mass.

\section{Conclusions}
We have determined the spectrum of the lowest lying baryon states for several
ensembles with two different gauge actions. The QCDOC machines and the RHMC algorithm
have made $2+1$ flavours of DWF ensembles possible for the first time. With limited
statistics, and only two different sea quark masses, we qualitatively reproduce the
experimental spectrum. There is limited evidence that finite size effects may be 
influencing the ground state baryons on the smaller volumes. Both the
Edinburgh plot, and the scaling analysis suggest that a programme of baryon
physics on larger volumes and at lighter quark mass will yield very interesting
results.

\section{Acknowledgements}
We thank Saul Cohen, Sam Li and Meifeng Lin for help generating the datasets used
in this work. We thank Dong Chen, Norman Christ, Saul Cohen, Calin
Cristian, Zhihua Dong, Alan Gara, Andrew Jackson, Chulwoo Jung,
Changhoan Kim, Ludmila Levkova, Xiaodong Liao, Guofeng Liu, Robert
Mawhinney, Shigemi Ohta, Konstantin Petrov and Tilo Wettig for
developing with us the QCDOC machine and its software. This development
 and the resulting computer equipment used in this calculation were funded
 by the U.S. DOE grant DE-FG02-92ER40699, PPARC JIF grant PPA/J/S/1998/00756
 and by RIKEN. This work was supported by PPARC grant PPA/G/O/2002/00465.

%\bibliographystyle{JHEP-2}
%\bibliography{refs}

\begin{thebibliography}{1}

\bibitem{Furman:1994ky}
V.~Furman and Y.~Shamir, {\it Axial symmetries in lattice qcd with kaplan
  fermions},  {\em Nucl. Phys.} {\bf B439} (1995) 54--78
  [\href{http://arXiv.org/abs/hep-lat/9405004}{{\tt hep-lat/9405004}}].
%%CITATION = HEP-LAT 9405004;%%

\bibitem{Vranas:1997da}
P.~M. Vranas, {\it Chiral symmetry restoration in the schwinger model with
  domain wall fermions},  {\em Phys. Rev.} {\bf D57} (1998) 1415--1432
  [\href{http://arXiv.org/abs/hep-lat/9705023}{{\tt hep-lat/9705023}}].
%%CITATION = HEP-LAT 9705023;%%

\bibitem{deForcrand:1999bi}
{\bf QCD-TARO} Collaboration, P.~de~Forcrand {\em et.~al.}, {\it
  Renormalization group flow of su(3) lattice gauge theory: Numerical studies
  in a two coupling space},  {\em Nucl. Phys.} {\bf B577} (2000) 263--278
  [\href{http://arXiv.org/abs/hep-lat/9911033}{{\tt hep-lat/9911033}}].
%%CITATION = HEP-LAT 9911033;%%

\bibitem{Iwasaki:1984cj}
Y.~Iwasaki and T.~Yoshie, {\it Renormalization group improved action for su(3)
  lattice gauge theory and the string tension},  {\em Phys. Lett.} {\bf B143}
  (1984) 449.
%%CITATION = PHLTA,B143,449;%%

\bibitem{Blum:2000kn}
T.~Blum {\em et.~al.}, {\it Quenched lattice qcd with domain wall fermions and
  the chiral limit},  {\em Phys. Rev.} {\bf D69} (2004) 074502
  [\href{http://arXiv.org/abs/hep-lat/0007038}{{\tt hep-lat/0007038}}].
%%CITATION = HEP-LAT 0007038;%%

\bibitem{PeterBoyle}
P.~Boyle. \emph{Localisation and chiral symmetry in 2+1 flavour domain wall
  QCD}, in proceedings of \emph{XXIIIrd International Symposium on Lattice
  Field Theory}, PoS(LAT2005)141.

\bibitem{Kennedy:1998cu}
A.~D. Kennedy, I.~Horvath and S.~Sint, {\it A new exact method for dynamical
  fermion computations with non-local actions},  {\em Nucl. Phys. Proc. Suppl.}
  {\bf 73} (1999) 834--836 [\href{http://arXiv.org/abs/hep-lat/9809092}{{\tt
  hep-lat/9809092}}].
%%CITATION = HEP-LAT 9809092;%%

\bibitem{MikeClark}
M.~Clark. \emph{Algorithm Shootout: R versus RHMC}, in proceedings of
  \emph{XXIIIrd International Symposium on Lattice Field Theory},
  PoS(LAT2005)115.

\bibitem{KoichiHashimoto}
K.~Hashimoto, T.~Izubuchi and J.~Noaki. \emph{The static quark potential in 2+1
  flavour Domain Wall QCD from QCDOC}, in proceedings of \emph{XXIIIrd
  International Symposium on Lattice Field Theory}, PoS(LAT2005)093.

\end{thebibliography}

\providecommand{\href}[2]{#2}\begingroup\raggedright\endgroup

\end{document}